\documentclass[showpacs,aps]{revtex4}
\usepackage{amsmath}
\usepackage{times}
\usepackage{amssymb}
\usepackage{graphicx}
\usepackage{float}
\usepackage{color}
\usepackage[utf8]{inputenc}
\def\<{\langle}
\def\>{\rangle}

\begin{document}
\today
\title{
Fluctuations of the heat flux of a one-dimensional 
hard particle gas
}
 \author{\'Eric Brunet, Bernard Derrida and Antoine Gerschenfeld}
\affiliation{ Laboratoire de Physique Statistique, Ecole Normale
Sup\'erieure, UPMC Paris 6,Universit\'e Paris Diderot, CNRS,
24 rue Lhomond, 75231 Paris Cedex 05 - France}

\begin{abstract}
Momentum-conserving one-dimensional models are known to exhibit anomalous
Fourier's law, with a thermal conductivity varying as a power law of the
system size. Here we measure, by numerical simulations, several cumulants of
the heat flux of a one-dimensional hard particle gas. We find that the
cumulants, like the conductivity, vary as power laws of the system size. Our
results also indicate that cumulants higher than the second follow
different power laws when one compares the ring geometry at equilibrium and
the linear case in contact with two heat baths (at equal or unequal
temperatures).

keywords: current fluctuations, anomalous Fourier law, hard particle gas
\end{abstract}


\maketitle

\date{\today}

\section*{Introduction}

Understanding the fluctuations of the flux of heat or of the current of
particles through systems in their steady state is a central question in non
equilibrium statistical mechanics. Since the discovery of the fluctuation
theorem \cite{ECM,GC,GC1,ES,Maes,Harris-S}, one knows that the probability distribution of these fluctuations has
some symmetry properties related to time-reversal symmetry. In most 
cases, the calculation of this distribution for a given microscopic model 
remains, however, a challenging issue.

\begin{figure}[h]
  \hspace{1cm}\includegraphics[width=10cm]{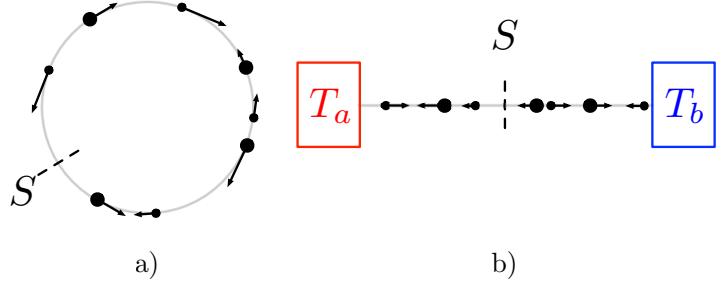}
  \caption{We measure the cumulants of the integrated current $Q_t$ of a
  one-dimensional hard particle gas with alternating masses $1$ and $m_2>1$:
  fig. \ref{geom}.a) for an even number of particles on a ring;
  fig.\;\ref{geom}.b) for an odd number of particles in contact with two heat
  baths (open system). In our simulation, we measure the flux $Q_t$ through a
  section $S$, located anywhere on the ring in fig.\;\ref{geom}.a) and
  half-way between the two heat baths in fig.\;\ref{geom}.b); we also measure
  the flux $\overline Q_t$ averaged over the whole system.}
  \label{geom}
\end{figure}

Over the last few years, several exact expressions for the distribution of
these fluctuations have nevertheless been obtained for diffusive systems, such
as lattice gases \cite{BD,BDGJL5,BDGJL6,ADLW,HG,Imparato,HRS1,derrida2007}
(for instance the one-dimensional symmetric simple exclusion process). For
diffusive systems in their steady state, in the two geometries of
fig.\;\ref{geom} (the ring geometry in equilibrium at temperature $T$ and the
open system, i.e.\;a finite system in contact with two heat baths at
temperatures $T_a$ and $T_b$), the cumulants $ \< Q_t^n \>_c$ of the flux of
energy $Q_t$ during a long time $t$ take, for a large system size $L$, the
following form:

\begin{equation}
  \lim_{t \to \infty}{1 \over t} \< Q_t^2 \>^{\rm ring} \simeq {1 \over L}
  A(T)\;;\ \ \ \lim_{t \to \infty}{1 \over t} \< Q_t^{2 n} \>_c^{\rm
  ring} \simeq {1 \over L^2} B_n(T) \ \ \ {\rm for } \ n \geq 2 \;;
  \ \ \ \lim_{t \to \infty}{1 \over t} \< Q_t^n \>_c^{\rm open} \simeq {1 \over L}
  C_n(T_a,T_b) \;.
  \label{dif}
\end{equation}

For the ring geometry (fig.\;\ref{geom}.a), explicit expressions of the
prefactors $A(T)$ and $ B_n(T)$ have been obtained for the symmetric simple
exclusion process, as well as for generic diffusive systems with one conserved
quantity \cite{ADLW}. For the open case (fig.\;\ref{geom}.b), the amplitudes
$C_n(T_a,T_b)$ are also known for the one-dimensional symmetric exclusion
process \cite{BD,BDGJL5,BDGJL6} (with expressions identical to those which had
been previously determined for disordered one-dimensional conductors
\cite{LLY,JSP}); they are also known for generic diffusive systems with one
conserved quantity \cite{HG,Imparato,HRS1,derrida2007,KMP,DDR,PM2}.

The $1/L$ dependence of $\< Q_t \>^{\rm open} /t$ for the open case means that
diffusive systems satisfy Fourier's law \cite{BLR}. Together with the
fluctuation-dissipation theorem, Fourier's law also implies the $1/L$ decay of
$\< Q_t^2\>^{\rm ring}/t$ (or of $\< Q_t^2\>^{\rm open}/t$ for $T_a=T_b$).

It is remarkable that the $L$ dependence (\ref{dif}) of {\it all} the
cumulants is generic for diffusive systems, with a few exceptions where some
prefactors may vanish and the decay with $L$ is faster \cite{ADLW}. All these
$L$ dependencies follow from the fact that, for large diffusive systems, local
equilibrium holds and can be treated within the {\it macrosopic fluctuation
theory} (MFT), a theory of diffusive systems which allows one to calculate
explicitly a number of properties of non-equilibrium steady states
\cite{KOV,HS,BDGJL1,BDGJL2,BDGJLY}. For systems with more than one conserved
quantity, much less has been done so far concerning the cumulants of the
current but, as long as the system is diffusive, one expects, from the MFT,
the same $L$ dependencies of the cumulants as in (\ref{dif}).

Momentum-conserving systems in one dimension are known to exhibit
anomalous Fourier's law \cite{LLP,LLP2,Dhar,PC,LW,BBO,MDN}, with an average current varying as a non
trivial power law of the system size $L$:
\begin{equation}
\lim_{t \to \infty}{1 \over t} \< Q_t \>^{\rm open} \sim L^{\alpha-1}
C_1(T_a,T_b) \;.
\label{mech-open}
\end{equation}
The exponent $\alpha$ is not easy to determine \cite{CP,LLP3,GNY,Dhar2,DN}. It
seems to vary with the systems studied; even for a given system, numerical
simulations or theoretical approaches do not always agree. The current
consensus is that several universality classes exist depending on the nature
of the non-linearity of the forces between the atoms \cite{NR,BLLLOP}.

As major numerical efforts have been already done to determine the exponent
$\alpha$, our goal here is to present numerical simulations, not to determine
$\alpha$ more accurately, but rather to look at the $L$ dependence of higher
cumulants of $Q_t$.

\section{Measurement of the cumulants for the hard particle gas}

The system we have decided to simulate is a one-dimensional gas
\cite{GNY,CP,DeN} of point particles with hard core interactions. It was
mainly chosen for its simple dynamics: masses follow ballistic motions between
successive collisions, which are elastic. If the particles were all of equal
masses, the velocities of colliding particles would simply be exchanged, and
the transport of energy would be the same as for an ideal gas; this is why, as
in previous studies of hard particle gases, we chose here a two-mass system,
with alternating particles of masses 1 and $m_2$.

For the ring geometry, we consider an even number $N$ of masses on a circle of
length $L=N$, initially in microcanonical equilibrium at fixed energy $E=N$
and zero total momentum: the total energy and the momentum remain of course
conserved by the dynamics.

For the open case in contact with two heat baths at unequal (or equal)
temperatures, we take an odd number $N$ of particles in a one-dimensional box
of size $L=N$, and we choose the particles closest to the boundaries to be of
mass 1. The heat baths at these boundaries are implemented in the following
way: whenever a particle hits a boundary, it is reflected as if a thermalized
particle was entering the system from the bath, so that the total number of
particles in the system remains $N$.

As the steady state is in general not known in the open case, we started our
first sample with an initial condition chosen at random; then, the initial
configuration of each new sample was taken to be the final configuration of
the previous sample. Therefore, apart from transient effects affecting the
first samples (which represent a very small fraction of the total), the
initial configurations of our samples are typical of steady state
configurations.

\begin{figure}[H]
  \includegraphics{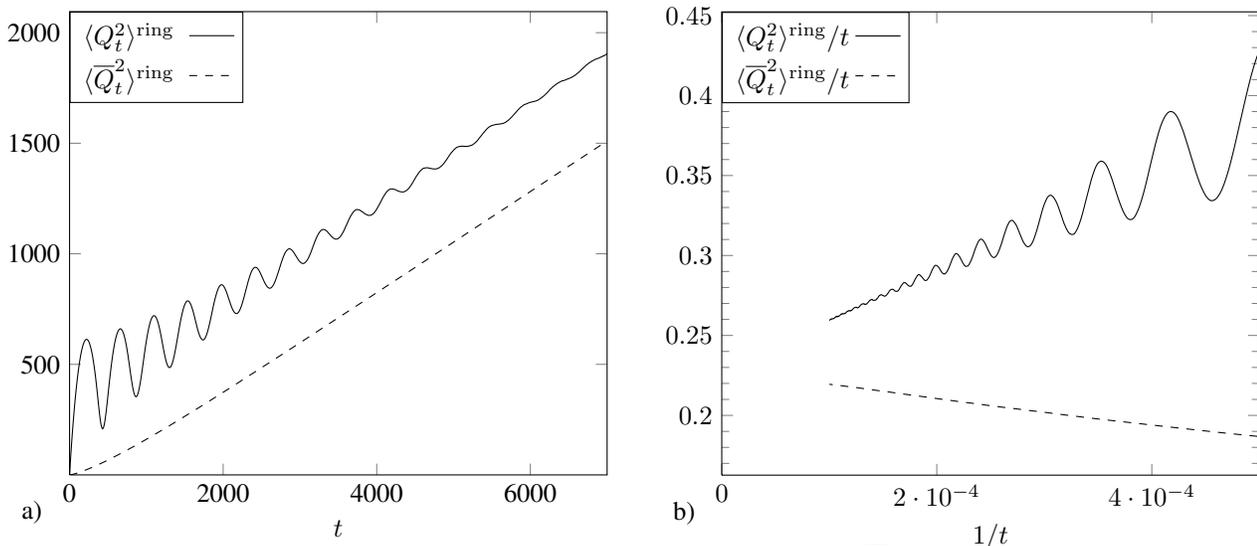}\vspace{-0.5cm}
  \caption{Second cumulant of the energy flux $Q_t$ through a section and of
  its space average $\overline Q_t$ on a ring of $N=800$ particles 
  (fig.\;\ref{fig:fits}.a). When plotted as functions of $1/t$ 
  (fig.\;\ref{fig:fits}.b), the two ratios $\< Q_t^2 \> / t$ converge, when
  $t\to\infty$, to a common value $\simeq0.23$.}
\label{fig:fits}
\end{figure}

Through any section of our system, the flux of energy $Q_t$ is the algebraic
sum of the kinetic energies of the particles crossing the section $S$ during a
time interval $t$. In the steady state, the statistical properties of $Q_t$
depend on where the section is located (at least in the open geometry). On
the other hand, one expects (under the assumption that the internal energy
cannot grow indefinitely) that the long time limit of the ratios $\< Q_t^n
\>_c/t$ are independent of where $Q_t$ is measured. In our simulations, we
measure for the same samples the flux of energy $Q_t$ through a fixed section
$S$ located at position $L/2$ in the open geometry (and anywhere on the ring
geometry) and its integrated value $\overline Q_t$ averaged over the whole
system.

As shown in fig.\;\ref{fig:fits}.a), the cumulants obtained from these two
measurements behave differently at finite time, but both exhibit a linear
growth for large $t$. Fig.\;\ref{fig:fits}.b) shows that, when the ratios $\<
Q_t^2 \>_c/t$ are plotted versus $1/t$, the two sets of data converge to a
common value in the long time limit. This was the case for all the cumulants
we were able to measure: for all the results shown below, the procedure of
fig.\;\ref{fig:fits}.b) was used to estimate the asymptotic values of the
cumulants $\<Q_t^n\>_c /t$.

Note that the $1/t$ convergence of fig.\;2.b) can easily be understood : if
the correlation function $\< J(t_1) J(t_2)\>_c = f(t_2-t_1)$ of the energy
current $J(t) = \partial_t Q_t$ decays fast enough in the steady state or at
equilibrium, then $\< Q_t^2\>_c = \iint_0^t \text{d}t_1 \text{d}t_2 \< J(t_1)
J(t_2)\>_c = 2t\int_0^t f(\tau) \text{d}\tau - 2 \int_0^t \tau f(\tau)
\text{d}\tau$ becomes of the form $A t + B$ in the large time limit, so that
$\< Q_t^2\>_c /t \sim A + B/t$.

\ \\
{\it Remark:} all the cumulants we could measure grow linearly with time for
large $t$. On the ring geometry, we only observed this growth when performing
microcanonical sampling at fixed total energy $E$ and momentum $P$ (we took
$E=N$ and $P=0$). When $E$ is allowed to fluctuate (canonical ensemble) while
keeping $P=0$, the cumulants exhibit a faster growth ($\< Q_t^{2n}\>\sim
t^n$). This is due to the conservation of $E$ by the dynamics on the
ring, which introduces non decaying current-current correlations in the
long time limit.

\section{Size dependence of the cumulants}
\begin{figure}[h]
  \centerline{\includegraphics{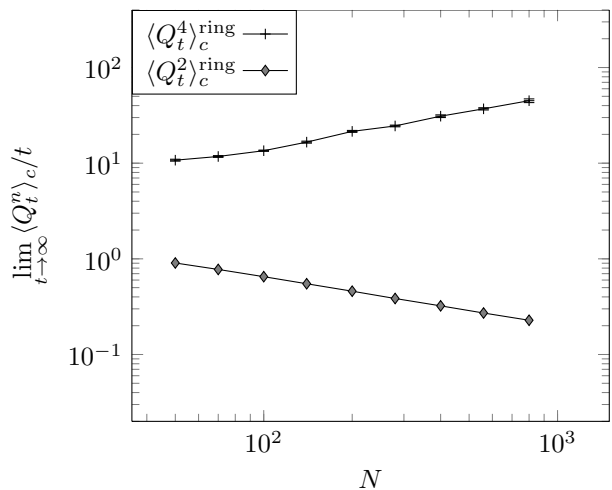}}\vspace{-3mm}
  \caption{Asymptotic values of $\< Q_t^n \>_c / t$ on a ring of $50 \leq N
  \leq 800$ particles of alternating masses 1 and 2.62, with total energy $N$
  and total momentum $0$. While the second cumulant decreases as 
  $N^{\alpha-1}$, with $\alpha\simeq 0.5$, the fourth cumulant increases with
   $N$.}
  \label{fig:ring}
\end{figure}

\begin{figure}[h]
  \includegraphics{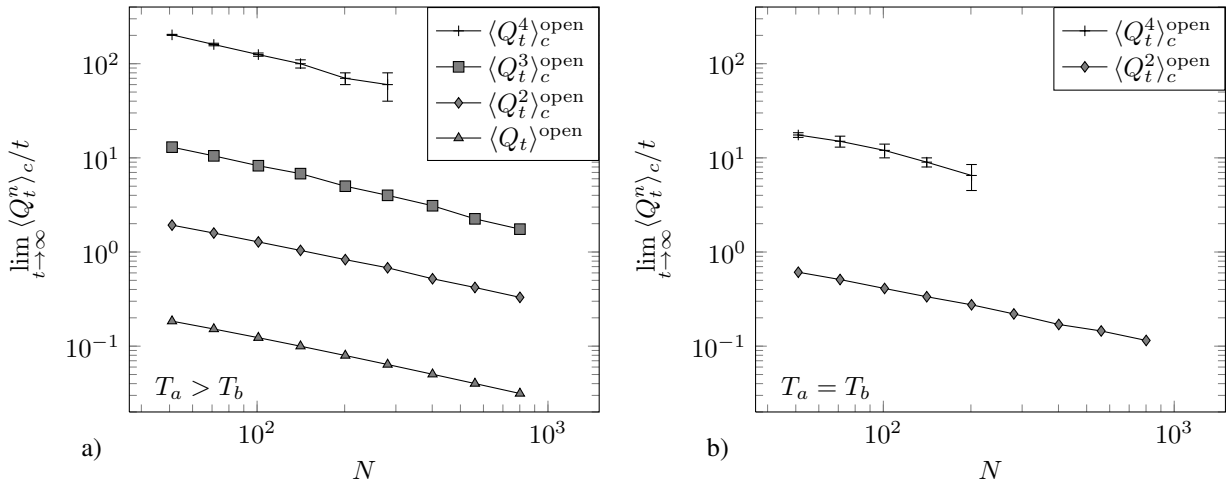}\vspace{-3mm}
  \caption{Asymptotic values of $\< Q_t^n \>_c / t$ on an open system with $51
  \leq N \leq 801$ particles, with alternating masses 1 and 2.62, between heat
  baths at temperatures $T_a = 2$ and $T_b = 1$ (fig.\;\ref{fig:open}.a) and
  $T_a = T_b = 1$ (fig.\;\ref{fig:open}.b). All cumulants seem to decrease as
  $N^{\alpha-1}$, with $0.25\leq \alpha \leq 0.4$.}
  \label{fig:open}
\end{figure}

Fig.\;\ref{fig:ring} and \ref{fig:open} show the asymptotic values of $\<
Q_t^n\>_c/t$ we obtained for $1\leq n \leq 4$. The cumulants were calculated
by averaging over a number of samples varying from $2 \cdot 10^8$ for $N=50$
to $2\cdot10^6$ for $N=800$, for the following systems:
\begin{itemize}
  \item a ring of $50\leq N \leq 800$ particles with total kinetic energy $N$
  and total momentum $0$;
  \item an open system of $51\leq N \leq 801$ particles between two heat baths
  at temperatures $T_a = 2$ and $T_b = 1$;
  \item the same open system between two heat baths at the same temperature
  $T_a = T_b = 1$.
\end{itemize}
In all cases, the particle density $N/L$ was exactly $1$. The mass of the
heavier particles, $m_2$ was taken to be $2.62$ (as in \cite{CP}); when we
repeated some of our simulations with a different mass ratio,
$m_2=(1+\sqrt{5})/2$ as in \cite{GNY}, we obtained
qualitatively similar results (not shown here).

The fourth cumulant of $Q_t$ is of course the hardest to obtain: we were only
able to measure the asymptotic value of $\< Q_t^4 \>_c / t$ accurately for
$N\leq 201$ in the open case.

For the ring geometry, while the second cumulant decays like a power law, the
fourth cumulant increases with the system size. Hence the picture is very
different from the diffusive case (\ref{dif}): in addition to the anomalous
Fourier's law, we observe an increase of the fourth cumulant instead of a
decay.

For the open case, the situation looks more similar to the diffusive case :
all the cumulants seem to decrease with comparable power laws of the system
size, albeit with a different exponent than in the diffusive case (\ref{dif}).

\section{Short-time behavior}

\begin{figure}[H]
  \centerline{\includegraphics{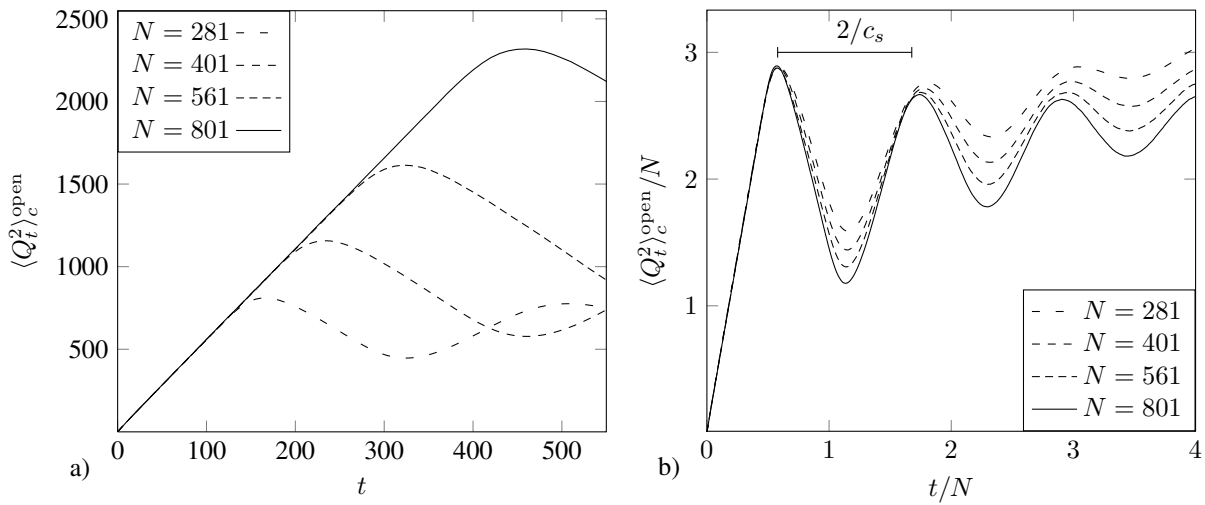}}\vspace{-3mm}
  \caption{Short-time behavior of $\< Q_t^2\>_c$ on an open system of $281
  \leq N \leq 801$ particles between heat baths at the same temperature $T_a =
  T_b = 1$. In a time range increasing with system size, the second cumulant
  behaves as on an infinite system at equilibrium (fig.\;\ref{fig:courts}.a);
  it then exhibits damped oscilllations on time scales of order $N$
  (fig.\;\ref{fig:courts}.b).}
  \label{fig:courts}
\end{figure}

In our simulations (fig.\;\ref{fig:fits}.a), we measure the whole time
dependence of the cumulants $\<Q_t^n\>_c$. As fig.\;\ref{fig:courts} shows,
the short-time behavior of $\<Q_t^n\>_c$ is independent of the system size $N$
over a time range which increases with $N$. This can be interpreted as the
fact that, over this time range, the system behaves as if it was infinite : in
turn, this allows us to study the behavior of the cumulants on an infinite
system. Fig.\;\ref{fig:courts}.a) shows that, for an infinite system at
equilibrium, $\<Q_t^2\>_c$ grows linearly with time; other results (not shown
here) indicate that $\<Q_t^4\>_c$ is also linear in time for the same system.
This is another major difference with infinite diffusive systems \cite{DG},
for which all the cumulants of $Q_t$ grow asymptotically as $\sqrt{t}$.

At intermediate times, all the $\< Q_t^n \>_c$ exhibit periodic oscillations
for $n\geq 2$, with a period proportional to system size (as can be seen for
the second cumulant on fig.\;\ref{fig:courts}.b): for the second cumulant,
they can be fitted by an exponentially damped sine function \cite{Bishop}. The periods can
be understood from the adiabatic sound velocity $c_s$ in the hard particle
gas, given by
$$c_s = \sqrt{\gamma P \over \rho}\;;$$
here, $\gamma = 3$ (for an one-dimensional monoatomic gas), $P= N k T / L=2$
(since the average energy, $E/N = kT/2$, is taken to be $1$), and $\rho =
(1+m_2) / 2$, so that $c_s = 1.82$ for $m_2=2.62$. For the open system, the
period is close to $2 N /c_s$ (shown on fig.\;\ref{fig:courts}.b), which is
the time for a sound wave originating in $S$ to come back to $S$ in the same
direction, having been reflected once on each boundary; for the ring, our data
(not shown here) exhibits a period close to $N/c_s$, the time for a sound wave
starting from $S$ to go around the ring.

\section{Conclusion}

In this letter, we have shown that, for a hard particle gas at equilibrium and
out of equilibrium, the cumulants of the flux of energy scale as power laws of
the system size. These power laws differ from the ones expected for diffusive
systems. Hence, our data shows that the anomalous Fourier's law of
momentum-conserving systems is also characteristic of higher cumulants of the
current. The difference between momentum-conserving and diffusive systems can
also be seen for an infinite system at equilibrium, in which $\<Q_t^2\>$ grows
as $t$ instead of $\sqrt{t}$.

More numerical work is certainly needed to accurately determine the exponents
characteristic of the cumulants of the energy flux, in particular to check
whether, in the open case, all the cumulants decay with the same power law;
one could try to compare the exponents seen for the ring and for the open
geometry.

We have focused here on the fluctuations of the flux of energy, which is
one of the conserved quantities of this hard particle gas. As momentum is also
conserved, it would be interesting to study the size dependence of the
fluctuations of the momentum flux in a similar way.

It would also be interesting to investigate whether similar power laws of the
cumulants can be observed in other momentum-conserving systems, such as
anharmonic chains like the Fermi-Pasta-Ulam models (which are known to exhibit
anomalous Fourier's law). The study of the cumulants could be a good test for
existing theoretical approaches \cite{MN,LuS}, such as the mode-coupling
theory \cite{PR,Lepri,DLRP,WL}.
or the Boltzmann-Langevin equation \cite{BZ,Perev}. In particular, one may
wonder whether, for momentum-conserving systems, there exists an analog of the
universal ratios of the cumulants which are expected for diffusive systems on
a ring \cite{ADLW}.

%\bibliographystyle{ieeetr}
%\bibliography{article3}

\begin{thebibliography}{10}

\bibitem{ECM}
D.~J. Evans, E.~G.~D. Cohen, and G.~P. Morriss, ``Probability of second law
  violations in shearing steady states,'' {\em Phys. Rev. Lett.}, vol.~71,
  p.~3616, Nov 1993.

\bibitem{GC}
G.~Gallavotti and E.~G.~D. Cohen, ``Dynamical ensembles in nonequilibrium
  statistical mechanics,'' {\em Phys. Rev. Lett.}, vol.~74, pp.~2694--2697, Apr
  1995.

\bibitem{GC1}
G.~Gallavotti and E.~Cohen, ``Dynamical ensembles in stationary states,'' {\em
  J. Stat. Phys.}, vol.~80, pp.~931--970, Sep 1995.

\bibitem{ES}
D.~J. Evans and D.~J. Searles, ``The fluctuation theorem,'' {\em Adv. Phys.},
  vol.~51, no.~7, pp.~1529--1585, 2002.

\bibitem{Maes}
C.~Maes, ``The fluctuation theorem as a gibbs property,'' {\em J. Stat. Phys.},
  vol.~95, pp.~367--392, Apr 1999.

\bibitem{Harris-S}
R.~J. Harris and G.~M. Schutz, ``Fluctuation theorems for stochastic
  dynamics,'' {\em J. Stat. Mech: Theory Exp.}, vol.~2007, no.~07, p.~P07020.

\bibitem{BD}
T.~Bodineau and B.~Derrida, ``Current fluctuations in nonequilibrium diffusive
  systems: An additivity principle,'' {\em Phys. Rev. Lett.}, vol.~92,
  p.~180601, May 2004.

\bibitem{BDGJL5}
L.~Bertini, A.~De~Sole, D.~Gabrielli, G.~Jona-Lasinio, and C.~Landim, ``Current
  fluctuations in stochastic lattice gases,'' {\em Phys. Rev. Lett.}, vol.~94,
  p.~030601, Jan 2005.

\bibitem{BDGJL6}
L.~Bertini, A.~Sole, D.~Gabrielli, G.~Jona-Lasinio, and C.~Landim, ``Non
  equilibrium current fluctuations in stochastic lattice gases,'' {\em J. Stat.
  Phys.}, vol.~123, pp.~237--276, Apr 2006.

\bibitem{ADLW}
C.~Appert-Rolland, B.~Derrida, V.~Lecomte, and F.~van Wijland, ``Universal
  cumulants of the current in diffusive systems on a ring,'' {\em Phys. Rev.
  E}, vol.~78, p.~021122, Aug 2008.

\bibitem{HG}
P.~I. Hurtado and P.~L. Garrido, ``Current fluctuations and statistics during a
  large deviation event in an exactly solvable transport model,'' {\em J. Stat.
  Mech: Theory Exp.}, vol.~2009, no.~02, p.~P02032.

\bibitem{Imparato}
A.~Imparato, V.~Lecomte, and F.~van Wijland, ``Equilibriumlike fluctuations in
  some boundary-driven open diffusive systems,'' {\em Phys. Rev. E}, vol.~80,
  p.~011131, Jul 2009.

\bibitem{HRS1}
R.~J. Harris, A.~Rakos, and G.~M. Schutz, ``Current fluctuations in the
  zero-range process with open boundaries,'' {\em J. Stat. Mech: Theory Exp.},
  vol.~2005, no.~08, p.~P08003.

\bibitem{derrida2007}
B.~Derrida, ``Non-equilibrium steady states: fluctuations and large deviations
  of the density and of the current,'' {\em J. Stat. Mech: Theory Exp.},
  vol.~2007, no.~07, p.~P07023.

\bibitem{LLY}
H.~Lee, L.~S. Levitov, and A.~Y. Yakovets, ``Universal statistics of transport
  in disordered conductors,'' {\em Phys. Rev. B}, vol.~51, pp.~4079--4083, Feb
  1995.

\bibitem{JSP}
A.~N. Jordan, E.~V. Sukhorukov, and S.~Pilgram, ``Fluctuation statistics in
  networks: A stochastic path integral approach,'' {\em Journal of Mathematical
  Physics}, vol.~45, no.~11, pp.~4386--4417, 2004.

\bibitem{KMP}
C.~Kipnis, C.~Marchioro, and E.~Presutti, ``Heat flow in an exactly solvable
  model,'' {\em J. Stat. Phys.}, vol.~27, pp.~65--74, Jan 1982.

\bibitem{DDR}
B.~Derrida, B.~Dou{\c c}ot, and P.-E. Roche, ``Current fluctuations in the
  one-dimensional symmetric exclusion process with open boundaries,'' {\em J.
  Stat. Phys.}, vol.~115, pp.~717--748, May 2004.

\bibitem{PM2}
S.~Prolhac and K.~Mallick, ``Cumulants of the current in a weakly asymmetric
  exclusion process,'' {\em J. Phys. A: Math. Theor.}, vol.~42, no.~17,
  p.~175001, 2009.

\bibitem{BLR}
F.~Bonetto, J.~L. Lebowitz, and L.~Rey-Bellet, ``Fourier's law: a challenge to
  theorists,'' in {\em Mathematical physics 2000}, pp.~128--150, London: Imp.
  Coll. Press, 2000.

\bibitem{KOV}
C.~Kipnis, S.~Olla, and S.~R.~S. Varadhan, ``Hydrodynamics and large deviation
  for simple exclusion processes,'' {\em Communications on Pure and Applied
  Mathematics}, vol.~42, no.~2, pp.~115--137, 1989.

\bibitem{HS}
H.~Spohn, {\em Large Scale Dynamics of Interacting Particles}.
\newblock Texts and Monographs in Physics, Springer, November 1991.

\bibitem{BDGJL1}
L.~Bertini, A.~De~Sole, D.~Gabrielli, G.~Jona-Lasinio, and C.~Landim,
  ``Fluctuations in stationary nonequilibrium states of irreversible
  processes,'' {\em Phys. Rev. Lett.}, vol.~87, p.~040601, Jul 2001.

\bibitem{BDGJL2}
L.~Bertini, A.~De~Sole, D.~Gabrielli, G.~Jona-Lasinio, and C.~Landim,
  ``Macroscopic fluctuation theory for stationary non-equilibrium states,''
  {\em J. Stat. Phys.}, vol.~107, pp.~635--675, May 2002.

\bibitem{BDGJLY}
L.~Bertini, A.~De~Sole, D.~Gabrielli, G.~Jona-Lasinio, and C.~Landim, ``Large
  deviation approach to non equilibrium processes in stochastic lattice
  gases,'' {\em Bulletin of the Brazilian Mathematical Society}, vol.~37,
  pp.~611--643, Dec 2006.

\bibitem{LLP}
S.~Lepri, R.~Livi, and A.~Politi, ``Heat conduction in chains of nonlinear
  oscillators,'' {\em Phys. Rev. Lett.}, vol.~78, pp.~1896--1899, Mar 1997.

\bibitem{LLP2}
S.~Lepri, R.~Livi, and A.~Politi, ``Thermal conduction in classical
  low-dimensional lattices,'' {\em Phys. Rep.}, vol.~377, pp.~1--80(80), April
  2003.

\bibitem{Dhar}
A.~Dhar, ``Heat transport in low-dimensional systems,'' {\em Adv. Phys.},
  vol.~57, no.~5, pp.~457--537, 2008.

\bibitem{PC}
T.~c.~v. Prosen and D.~K. Campbell, ``Momentum conservation implies anomalous
  energy transport in 1d classical lattices,'' {\em Phys. Rev. Lett.}, vol.~84,
  pp.~2857--2860, Mar 2000.

\bibitem{LW}
B.~Li and J.~Wang, ``Anomalous heat conduction and anomalous diffusion in
  one-dimensional systems,'' {\em Phys. Rev. Lett.}, vol.~91, p.~044301, Jul
  2003.

\bibitem{BBO}
G.~Basile, C.~Bernardin, and S.~Olla, ``Momentum conserving model with
  anomalous thermal conductivity in low dimensional systems,'' {\em Phys. Rev.
  Lett.}, vol.~96, p.~204303, May 2006.

\bibitem{MDN}
T.~Mai, A.~Dhar, and O.~Narayan, ``Equilibration and universal heat conduction
  in fermi-pasta-ulam chains,'' {\em Phys. Rev. Lett.}, vol.~98, p.~184301, May
  2007.

\bibitem{CP}
G.~Casati and T.~c.~v. Prosen, ``Anomalous heat conduction in a one-dimensional
  ideal gas,'' {\em Phys. Rev. E}, vol.~67, p.~015203, Jan 2003.

\bibitem{LLP3}
S.~Lepri, R.~Livi, and A.~Politi, ``Universality of anomalous one-dimensional
  heat conductivity,'' {\em Phys. Rev. E}, vol.~68, p.~067102, Dec 2003.

\bibitem{GNY}
P.~Grassberger, W.~Nadler, and L.~Yang, ``Heat conduction and entropy
  production in a one-dimensional hard-particle gas,'' {\em Phys. Rev. Lett.},
  vol.~89, p.~180601, Oct 2002.

\bibitem{Dhar2}
A.~Dhar, ``Heat conduction in a one-dimensional gas of elastically colliding
  particles of unequal masses,'' {\em Phys. Rev. Lett.}, vol.~86,
  pp.~3554--3557, Apr 2001.

\bibitem{DN}
A.~Dhar and O.~Narayan, ``Dhar et al. reply:,'' {\em Phys. Rev. Lett.},
  vol.~100, p.~199402, May 2008.

\bibitem{NR}
O.~Narayan and S.~Ramaswamy, ``Anomalous heat conduction in one-dimensional
  momentum-conserving systems,'' {\em Phys. Rev. Lett.}, vol.~89, p.~200601,
  Oct 2002.

\bibitem{BLLLOP}
G.~Basile, L.~Delfini, S.~Lepri, R.~Livi, S.~Olla, and A.~Politi, ``Anomalous
  transport and relaxation in classical one-dimensional models,'' {\em Eur.
  Phys. J. Special Topics}, vol.~151, pp.~85--93, dec 2007.

\bibitem{DeN}
J.~M. Deutsch and O.~Narayan, ``One-dimensional heat conductivity exponent from
  a random collision model,'' {\em Phys. Rev. E}, vol.~68, p.~010201, Jul 2003.

\bibitem{DG}
B.~Derrida and A.~Gerschenfeld, ``Current fluctuations in one dimensional
  diffusive systems with a step initial density profile,'' {\em J. Stat.
  Phys.}, vol.~137, pp.~978--1000, Dec 2009.

\bibitem{Bishop}
M.~Bishop, ``Collective modes of one-dimensional {L}ennard-{J}ones systems,''
  {\em J. Stat. Phys.}, vol.~29, pp.~623--629, Nov 1982.

\bibitem{MN}
T.~Mai and O.~Narayan, ``Universality of one-dimensional heat conductivity,''
  {\em Phys. Rev. E}, vol.~73, p.~061202, Jun 2006.

\bibitem{LuS}
J.~Lukkarinen and H.~Spohn, ``Anomalous energy transport in the {FPU}-{$\beta$}
  chain,'' {\em Comm. Pure Appl. Math.}, vol.~61, no.~12, pp.~1753--1786, 2008.

\bibitem{PR}
Y.~Pomeau and P.~R{\'e}sibois, ``Time dependent correlation functions and
  mode-mode coupling theories,'' {\em Phys. Rep.}, vol.~19, no.~2, pp.~63 --
  139, 1975.

\bibitem{Lepri}
S.~Lepri, ``Relaxation of classical many-body hamiltonians in one dimension,''
  {\em Phys. Rev. E}, vol.~58, pp.~7165--7171, Dec 1998.

\bibitem{DLRP}
L.~Delfini, S.~Lepri, R.~Livi, and A.~Politi, ``Anomalous kinetics and
  transport from 1d self-consistent mode-coupling theory,'' {\em J. Stat. Mech:
  Theory Exp.}, vol.~2007, no.~02, p.~P02007.

\bibitem{WL}
J.-S. Wang and B.~Li, ``Mode-coupling theory and molecular dynamics simulation
  for heat conduction in a chain with transverse motions,'' {\em Phys. Rev. E},
  vol.~70, p.~021204, Aug 2004.

\bibitem{BZ}
M.~Bixon and R.~Zwanzig, ``{B}oltzmann-{L}angevin equation and hydrodynamic
  fluctuations,'' {\em Phys. Rev.}, vol.~187, pp.~267--272, Nov 1969.

\bibitem{Perev}
A.~Pereverzev, ``{F}ermi-{P}asta-{U}lam $\beta$ lattice: {P}eierls equation and
  anomalous heat conductivity,'' {\em Phys. Rev. E}, vol.~68, p.~056124, Nov
  2003.

\end{thebibliography}

\end{document}